\documentclass[aps,pre,twocolumn,superscriptaddress]{revtex4-1}

\usepackage{graphicx}
\usepackage{dcolumn}
\usepackage{bm}
\usepackage{amsmath}
\usepackage{amsfonts}
\usepackage{amssymb}
\usepackage{graphicx}
\usepackage{float}
\usepackage{color}
\newcommand{\rev}{\textcolor{black}}

\begin{document}


\title{Shear-induced contact area anisotropy explained by a fracture
mechanics model}

\author{A. Papangelo}
\affiliation{Dipartimento di Meccanica, Matematica e Management, Politecnico di Bari, Viale Japigia 182, 70126 Bari, Italy}
\affiliation{Hamburg University of Technology, Department of Mechanical Engineering, Am Schwarzenberg-Campus 1, 21073 Hamburg, Germany}

\author{J. Scheibert}
\affiliation{Univ Lyon, Ecole Centrale de Lyon, ENISE, ENTPE, CNRS, Laboratoire de Tribologie et Dynamique des Syst\`emes LTDS, UMR 5513, F-69134, Ecully, France}

\author{R. Sahli}
\affiliation{Univ Lyon, Ecole Centrale de Lyon, ENISE, ENTPE, CNRS, Laboratoire de Tribologie et Dynamique des Syst\`emes LTDS, UMR 5513, F-69134, Ecully, France}

\author{G. Pallares}
\affiliation{Univ Lyon, Ecole Centrale de Lyon, ENISE, ENTPE, CNRS, Laboratoire de Tribologie et Dynamique des Syst\`emes LTDS, UMR 5513, F-69134, Ecully, France}
\affiliation{CESI, LINEACT, Zone A\'eroportuaire M\'editerran\'ee, 34130 Mauguio, France}

\author{M. Ciavarella}
\affiliation{Dipartimento di Meccanica, Matematica e Management, Politecnico di Bari, Viale Japigia 182, 70126 Bari, Italy}
\affiliation{Hamburg University of Technology, Department of Mechanical Engineering, Am Schwarzenberg-Campus 1, 21073 Hamburg, Germany}
\email{mciava@poliba.it}

%
%
%


\date{\today}

\begin{abstract}
This paper gives a theoretical analysis for the fundamental problem of
anisotropy induced by shear forces on an adhesive contact, discussing the experimental data of the companion Letter. We present a fracture mechanics model where two phenomenological mode-mixity functions are introduced to describe
the weak coupling between modes I and II or I and III, which changes the
effective toughness of the interface. The mode-mixity functions have been
interpolated using the data of a single experiment and then used to predict
the behaviour of the whole set of experimental observations. The model extends an idea by Johnson and Greenwood, i.e. to solve purely mode I problems of adhesion in the presence of a non-axisymmetric Hertzian geometry, to the case of elliptical contacts sheared along their major or minor axis. Equality between the stress intensity factors and their critical values is imposed solely at the major and minor axes. We successfully validate our model against experimental data. The model predicts that the
punch geometry will affect both the shape and the overall decay of the sheared contact area.
\begin{description}
\item[PACS numbers]
81.40.Pq, 68.35.Np, 62.20.Qp
\end{description}
\end{abstract}

\pacs{Valid PACS appear here}
\maketitle


\section{Introduction}

The interplay of adhesion and friction is a problem of fundamental importance
in tribology, which ideally should be solved at all scales from tectonic
plates to atomic scales (for a recent review of multiscale methods and
problems in tribology, see \cite{Vakis}). In the particular case of soft
materials, it is already relatively well understood and plays a substantial role in
Nature: in many insects, for example, an equivalent of an "adhesive Coulomb
friction law" has been described, whereby the normal force to detach the
adhesive "pads" is proportional to the shear force simultaneously applied
(\cite{Autumn}, \cite{Labonte}, \cite{Gravish}). For soft materials, a finite
contact area is observed also under zero force due to adhesion \cite{JKR} and
as a consequence, friction is measured also under vanishing or even negative
normal forces \cite{Homola}\cite{Yoshua}. There is no unique framework to study
this interaction \cite{Mergel}: for instance for hard materials, although no macroscopic adhesion
is found and friction may have a number of origins, Rabinowicz \cite{rabino}%
\cite{Rabino1992} attempted to describe friction in terms of surface energy. Another example is the onset of sliding, for which fracture-like surface energy concepts have been used successfully \cite{Svet},\cite{pap2015},\cite{papa12015}.

Here, we consider typically soft materials, for which the first fracture
mechanics model and experiment for adhesion and friction interaction was
conceived for macroscopic smooth spheres by Savkoor \&\ Briggs \cite{Savkoor},
who extended the Johnson-Kendall-Roberts (JKR) model \cite{JKR} to the presence of tangential force. This
model however corresponded to a "purely brittle" model where the frictional
resistance was neglected and, as such, \textit{greatly underestimated} the interfacial toughness. In that respect, it has been observed that when mode
I combines with mode II or/and mode III (see Fig. \ref{setup}a), the interfacial toughness is greatly
increased. The physical explanations for this increase are various (e.g.
friction, plasticity, dislocation emission) and cannot be ascribed to a single
phenomenon \cite{Hutch1990}. Since then, a few phenomenological models have
been proposed (\cite{Joh1996},\cite{Johnson1997},\cite{WG2010}%
,\cite{Popov1},\cite{Filippov1},\cite{Ciavafacta},\cite{Pap2019}) which require a Mode-Mixity Function (MMF)
$f\left(  \psi\right)  $ \cite{Hutc1992} to describe the critical condition
for propagation%
\begin{equation}
G_{c}=G_{Ic}f\left(  \psi\right)
\end{equation}
where $G_{Ic}$ is mode I critical factor (or surface energy, if we assume
Griffith's concept), $G_{c}\ $is the critical energy release rate in mixed mode conditions and finally $\psi$ is the "phase angle"%
\begin{align}
\psi_2&=\arctan\left(  \frac{K_{II}}{K_{I}}\right)\\
\psi_3&=\arctan\left(  \frac{K_{III}}{K_{I}}\right)
\end{align}
being $K_{III}$, $K_{II}$ and $K_{I}$ respectively the mode III, mode II and mode I stress
intensity factors.

The most recent model in this field is perhaps that by Papangelo \&
Ciavarella \cite{Pap2019} who compared it with recent experimental measurements by
Mergel et al. \cite{Mergel}, and concluded that the transition to sliding is very
sensitive to the choice of the mode-mixity function. Papangelo \& Ciavarella's mode-mixity model \cite{Pap2019} suggests that upon shearing the contact can experience either a smooth transition from the JKR to the Hertzian contact area or an
unstable jump to the Hertzian solution where lighter normal forces favour the
latter behaviour. All Linear Elastic Fracture Mechanics (LEFM) models indicate a decay of the contact area with
force, but the overall evolution strongly depends on the effective form of the
MMF \cite{Pap2019}. Furthermore, the most up to date experimental evidences
show that for high normal forces, the decay of the contact area with the
tangential force is quadratic \cite{Sahli}, while for small normal forces
\cite{Mergel} it isn't. Experimental measurements of contact area evolution
show that the shape of contact area is circular, according to JKR theory, at
zero tangential force and shrinks in an elliptical-like fashion while the shear force
is increased (\cite{Mergel}, \cite{Sahli}, \cite{WG2010}). So far, all LEFM
models proposed (\cite{Savkoor},\cite{Joh1996},\cite{Johnson1997}%
,\cite{WG2010},\cite{Ciavafacta},\cite{Pap2019}) make the approximation to
consider the contact as circular, even when sheared.
This requires an averaging of the effects of mode II and mode III around the
periphery. However, it is well known that sphere/plane contacts loose their initial circularity when submitted to shear, indicating that axisymmetry is a very questionable assumption. Note that recent experimental investigations for rough interfaces composed of many asperities \cite{Sahli} have showed similar anisotropic real area reduction and morphology changes, as discussed extensively in the companion Letter \cite{Letter}. A better understanding of the simpler sphere/plane contacts is crucial to comprehend shear induced-anisotropy in rough contacts.

In the present paper, we shall extend the axisymmetric theory to include the
case of elliptical shrinking of single contact area with the shear force,
starting from either circular or even already elliptical contact area. Initial
ellipticity typically occurs in the case of rough contacts, where most
summits are mildly elliptical, the most common ratio of principal summit
curvatures being near 2:1 \cite{G2006}. 

The only assumption we make for simplicity is that either the major or the minor axis of the contact ellipse is
aligned with the shear force: results will show a sufficiently clear overall
picture. In the first part of the manuscript the theoretical model will be
introduced, while in the second part it will be validated against the
experimental results provided in the companion Letter \cite{Letter} and in Sahli et al. \cite{Sahli}.

\section{The approximate JKR theory for elliptical contacts}

In absence of tangential force, Johnson and Greenwood \cite{JG} (JG in the
following) developed an approximate JKR theory for adhesion of an Hertzian
profile with differing principal radii of curvature. The
contact problem is solved "approximately" in a sense that the equality of
the Stress Intensity Factor (SIF) to its critical value $K_{Ic}$ round the periphery is only satisfied at the
major and minor axis of the contact ellipse. JG assume a pressure distribution
equal to

\begin{equation}
p\left(  x,y\right)  =\frac{p_{1}-\alpha x^{2}-\beta y^{2}}{\sqrt{1-\left(
x/a\right)  ^{2}-\left(  y/b\right)  ^{2}}}\label{pJG}%
\end{equation}
where $a$ and $b$ are respectively the major and minor semi-axes of
the ellipse, $\left(  p_{1},\alpha,\beta\right)  $ are constants to be found
and $p\left(x,y\right)  $ is taken positive (negative) when compressive (tensile).
The stress intensity factors at the major and minor axis (respectively $a$
and $b$) are%
\begin{align}
K_{I}\left(  a\right)   &  =\left(  \alpha a^{2}-p_{1}\right)
\sqrt{\pi a}=K_{Ic}\label{Ka}\\
K_{I}\left(  b\right)   &  =\left(  \beta b^{2}-p_{1}\right)
\sqrt{\pi b}=K_{Ic}\label{Kb}%
\end{align}
JG impose the SIF at the major and minor axis to be equal to its
critical value which, by standard LEFM arguments, is
$K_{Ic}=\sqrt{2E^{\ast}G_{Ic}}$, where $E^{\ast}$ is the plane strain
composite modulus of the interface, and $G_{Ic}$ the mode I "toughness" or
surface energy$.$ Galin's \cite{Galin} theorem establishes that any pressure
distribution of the form (\ref{pJG}) produces a field of quadratic
displacements
\begin{equation}
w=w_{00}-w_{20}x^{2}-w_{02}y^{2}%
\end{equation}
where $w_{00}$ is the indentation and $\left(  w_{20},w_{02}\right)  $ are
constants to be found. Kalker \cite{Kalker} reveals the relation between the
sets of constants $\left(  \alpha,\beta\right)  $ and $\left(  w_{20}%
,w_{02}\right)  $
\begin{equation}
\left[
\begin{array}
[c]{c}%
w_{20}\\
w_{02}%
\end{array}
\right]  =\left(  \frac{b}{E^{\ast}}\right)  \left[
\begin{array}
[c]{c}%
\left(  \boldsymbol{D}+\boldsymbol{C}\right)  \alpha-\left(  b%
/a\right)  ^{2}\boldsymbol{C}\beta\\
-\boldsymbol{C}\alpha+\left\{  \boldsymbol{B+}\left(  b/a\right)
^{2}\boldsymbol{C}\right\}  \beta
\end{array}
\right]  =\left[
\begin{array}
[c]{c}%
1/2R_{1}\\
1/2R_{2}%
\end{array}
\right]  \label{comp}%
\end{equation}
where $\boldsymbol{K}\left(  e\right)  \boldsymbol{,E}\left(  e\right)
\boldsymbol{,B}\left(  e\right)  \boldsymbol{,C}\left(  e\right)
\boldsymbol{,D}\left(  e\right)  $ are complete elliptic integrals of argument
$e^{2}=1-g^{2}$ $\left(  g=b/a<1\right)  $ with $e^{2}\boldsymbol{D}%
\left(  e\right)  =\boldsymbol{K}\left(  e\right)  -\boldsymbol{E}\left(
e\right)  ,$ $\boldsymbol{B}\left(  e\right)  =\boldsymbol{K}\left(  e\right)
-\boldsymbol{D}\left(  e\right)  ,$ $e^{2}\boldsymbol{C}\left(  e\right)
=\boldsymbol{D}\left(  e\right)  -\boldsymbol{B}\left(  e\right)
$ and $\left(  R_{1},R_{2}\right)  $
are the principal radii of curvature. The problem is closed adding the
equation for the total normal force $P$%
\begin{equation}
P=2\pi ab\left[  p_{1}-\frac{1}{3}\left(  \alpha a^{2}+\beta
b^{2}\right)  \right]  \label{W}%
\end{equation}
or for the indentation $\delta$ $\left(  =w_{00}\right)  $ \cite{Kalker}%

\begin{equation}
\delta=\left(  \frac{b}{E^{\ast}}\right)  \left[  2p_{1}\boldsymbol{K}%
-\alpha a^{2}\boldsymbol{B}-\beta b^{2}\boldsymbol{D}\right]
\label{d}%
\end{equation}
which, in the original case of JG, closes the system of 5 equations
(\ref{Ka},\ref{Kb},\ref{comp},\ref{W} (or \ref{d}))\ in the 5 unknowns
$\left(  a,b,p_{1},\alpha,\beta\right)  $. For $R_{1}=R_{2}$ this
corresponds to the classical JKR solution.

\section{The effect of tangential force}

\subsection{Theoretical model}

\begin{figure}[ptb]
\begin{center}
\includegraphics[width=3.5in]{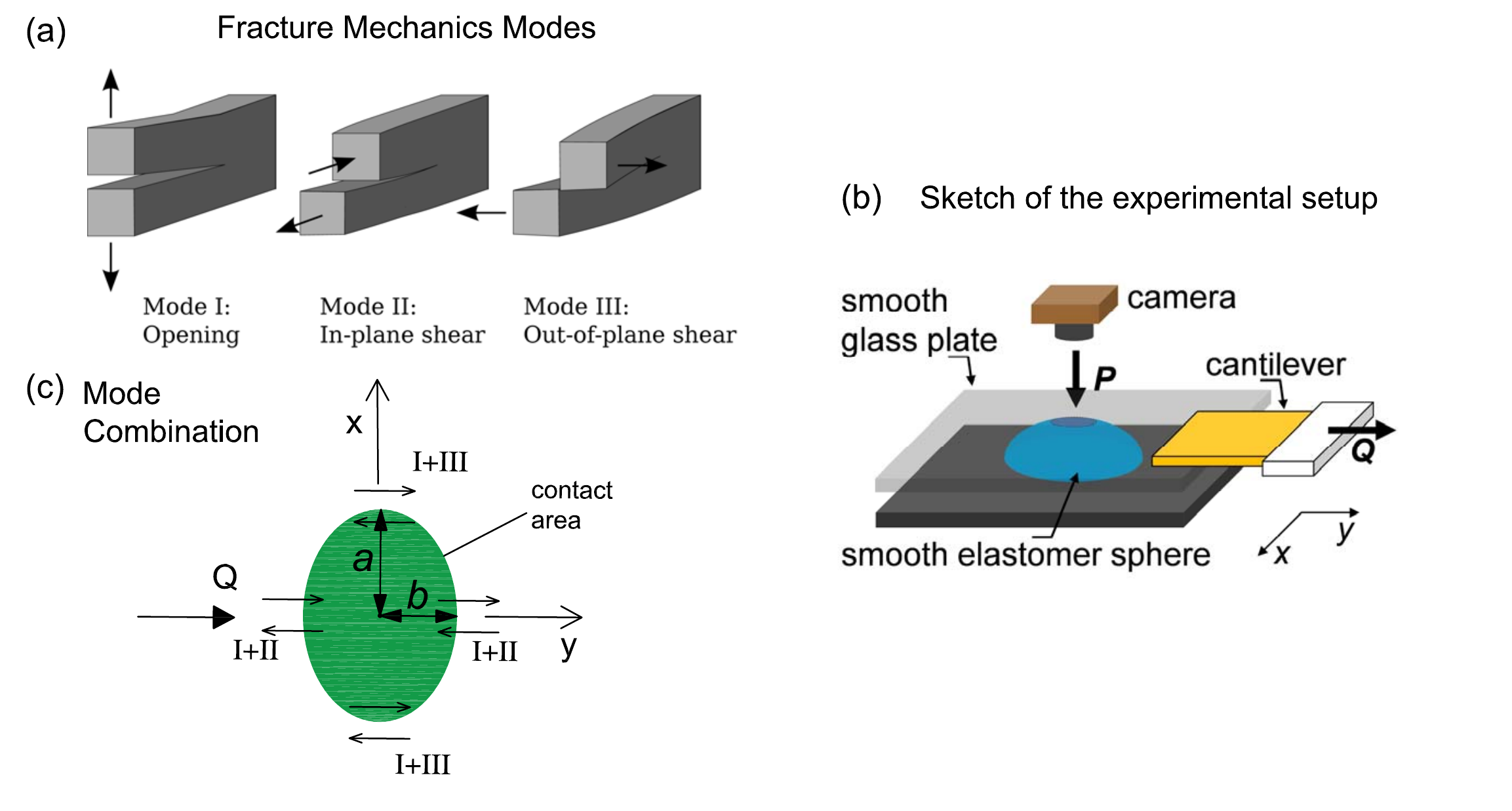}
\caption{(a) Fracture mechanics rupture modes \rev{(from \cite{wiki})}. (b) Sketch of the experimental setup used in Sahli et al. \cite{Sahli} and in the companion Letter \cite{Letter}. (c) Combination of the modes along the periphery of the contact patch.}
\label{setup}
\end{center}
\end{figure}

Assume that we have a sphere of radius $R$ in adhesive contact with a
halfspace (see Fig. \ref{setup}b).


If a tangential shearing force $Q$ is applied, and no slip occurs in the
contact area, a singular shear traction distribution of the form
\begin{equation}
q\left(  x,y\right)  =q_{0}/\sqrt{1-\left(  x/a\right)  ^{2}-\left(
y/b\right)  ^{2}}\label{q}%
\end{equation}
will arise at the interface. Experimental inspection of contact area in this
condition shows the contact patch is nearly elliptical and shrinks along the
direction of the applied shearing force (mode II), while remaining slightly
affected in the perpendicular direction (mode III) (see sketch Fig. \ref{setup}c, the companion Letter \cite{Letter},
\cite{Sahli}, \cite{Mergel} and \cite{WG2010}). A shear traction distribution of
the form (\ref{q}) gives a tangential force $Q=2\pi abq_{0}$ and
produces at the major axis $K_{II}\left(  a\right)  =0$ and $K_{III}%
\left(  a\right)  =q_{0}\sqrt{\pi a},$ while at the minor axis,
$K_{II}\left(  b\right)  =q_{0}\sqrt{\pi b},$ $K_{III}\left(
b\right)  =0$. The energy release rate according to standard Fracture
Mechanics arguments is $G=\frac{1}{2E^{\ast}}\left(  K_{I}^{2}+K_{II}%
^{2}+\frac{1}{1-\nu}K_{III}^{2}\right)  $ , thus using (\ref{Ka}%
,\ref{Kb})\ the equivalent SIF at the major "$a$" and minor "$b$" axes
are
\begin{align}
K_{eq}\left(  a\right)   &  =\sqrt{K_{I}^{2}\left(  a\right)
+\frac{1}{1-\nu}K_{III}^{2}\left(  a\right)  }=\sqrt{\left(  \alpha a^{2}%
-p_{1}\right)  ^{2}+2q_{0}^{2}}\sqrt{\pi a}\\
K_{eq}\left(  b\right)   &  =\sqrt{K_{I}^{2}\left(  b\right)
+K_{II}^{2}\left(  b\right)  }=\sqrt{\left(  \beta b^{2}-p_{1}\right)
^{2}+q_{0}^{2}}\sqrt{\pi b}%
\end{align}
The critical energy
needed for the external crack to advance, $G_{c}$, depends on the
"mode-mixity". Following Hutchinson \& Suo \cite{Hutc1992} we shall postulate
that $G_{c}$ depends on the phase angles $\psi_{2}=\arctan\left(  \frac
{K_{II}}{K_{I}}\right)  $ and $\psi_{3}=\arctan\left(  \frac{K_{III}}{K_{I}%
}\right)  $, thus at the minor (where we have modes I and II) and major (where we have modes I and III) axes we write respectively $G_{c}=G_{Ic}%
f_{II}\left(  \psi_{2}\right)  $ and $G_{c}=G_{Ic}f_{III}\left(  \psi
_{3}\right)  ,$ i.e.%

\begin{align}
\sqrt{\left(  \beta b^{2}-p_{1}\right)  ^{2}+q_{0}^{2}}\sqrt{\pi b} &
=K_{Ic}\sqrt{f_{II}\left(  \psi_{2}\right)  }\\
\sqrt{\left(  \alpha a^{2}-p_{1}\right)  ^{2}+\frac{1}{1-\nu}q_{0}^{2}}\sqrt{\pi a}
&  =K_{Ic}\sqrt{f_{III}\left(  \psi_{3}\right)  }%
\end{align}
where $f_{II}\left(  \psi_{2}\right)  $ and $f_{III}\left(  \psi_{3}\right)  $
are two MMFs which take into account the mixed-mode dependent toughness of the interface.

To sum up, the problem is reduced to a system of $5$ equations in the $5$
unknown $\left(  a,b,p_{1},\alpha,\beta\right)  $\footnote{The last
equation for the force can be replaced by the respective for indentation
(\ref{d}).}%

\begin{equation}
\left\{
\begin{array}
[c]{l}%
\sqrt{\left(  \beta b^{2}-p_{1}\right)  ^{2}+q_{0}^{2}}\sqrt{\pi b%
}-\sqrt{2E^{\ast}G_{Ic}f_{II}\left(  \psi_{2}\right)  }=0\\
\sqrt{\left(  \alpha a^{2}-p_{1}\right)  ^{2}+\frac{1}{1-\nu}q_{0}^{2}}\sqrt{\pi a%
}-\sqrt{2E^{\ast}G_{Ic}f_{III}\left(  \psi_{3}\right)  }=0\\
\left(  \frac{b}{E^{\ast}}\right)  \left[  \left(  \boldsymbol{D}%
+\boldsymbol{C}\right)  \alpha-\left(  b/a\right)  ^{2}\boldsymbol{C}%
\beta\right]  -\frac{1}{2R_{1}}=0\\
\left(  \frac{b}{E^{\ast}}\right)  \left[  -\boldsymbol{C}\alpha+\left\{
\boldsymbol{B+}\left(  b/a\right)  ^{2}\boldsymbol{C}\right\}
\beta\right]  -\frac{1}{2R_{2}}=0\\
P-2\pi ab\left[  p_{1}-\frac{1}{3}\left(  \alpha a^{2}+\beta
b^{2}\right)  \right]  =0
\end{array}
\right.  \label{system}%
\end{equation}
where, if the punch is axisymmetric\footnote{In deriving the model we start with the case of a sheared spherical punch. Nevertheless the model can be also used for non-axisymmetric Hertzian geometry provided that the tangential force is aligned with the minor or major axis.} $R_1=R_2=R$. In principle, if one knows how the interfacial toughness depends on the mode
combination, this problem can be solved exactly, with the sole approximation
that the equality of the SIFs with their critical values is guaranteed only at the major and minor
axes in line with JG approximation.

Next, the following dimensionless notation is introduced \cite{Maugis}%

\begin{equation}
\begin{aligned}
\gamma &  =\sqrt{\frac{R_{2}}{R_{1}}};\text{\quad\quad}R_{e}=\sqrt{R_{2}R_{1}%
};\text{\quad\quad}\xi=\left(  \frac{E^{\ast}R_{e}}{G_{Ic}}\right)
^{1/3};\\
\widetilde{a} &  =\frac{\xi a}{R_{e}};\text{\quad\quad}\widetilde{b} =\frac{\xi b}{R_{e}};\text{\quad\quad}
g  =\frac{b}{a};\text{\quad\quad
}\widetilde{\delta}=\frac{\xi^{2}\delta}{R_{e}}; \\
\widetilde{Q} & =\frac{Q}{R_{e}G_{Ic}};  \text{\quad\quad}\widetilde{P}=\frac{P}{R_{e}G_{Ic}
}; \text{\quad\quad}
\widetilde{\alpha}=\frac{R_{e}^{2}\alpha}{\xi E^{\ast}};\\
\widetilde{\beta} &  =\frac{R_{e}^{2}\beta}{\xi E^{\ast}};\text{\quad\quad
}\widetilde{p}_{1}=\frac{\xi p_{1}}{E^{\ast}};\text{\quad\quad}\widetilde
{q}_{0}=\frac{\xi q_{0}}{E^{\ast}}%
\end{aligned}
\end{equation}

\noindent{and} the system of eq. (\ref{system}) is written in dimensionless form

\begin{equation}
\left\{
\begin{array}
[c]{l}%
\sqrt{\left(  \widetilde{\beta}g^{2}\widetilde{a}^{2}-\widetilde{p}%
_{1}\right)  ^{2}+\left(  \frac{\widetilde{Q}}{2\pi\widetilde{a}^{2}%
g}\right)  ^{2}}\sqrt{\pi g\widetilde{a}}-\sqrt{2f_{II}\left(  \psi
_{2}\right)  }=0\\
\sqrt{\left(  \widetilde{\alpha}\widetilde{a}^{2}-\widetilde{p}%
_{1}\right)  ^{2}+\frac{1}{1-\nu}\left(  \frac{\widetilde{Q}}{2\pi\widetilde{a}^{2}%
g}\right)  ^{2}}\sqrt{\pi\widetilde{a}}-\sqrt{2f_{III}\left(  \psi
_{3}\right)  }=0\\
\widetilde{a}g\left[  \left(  \boldsymbol{D}+\boldsymbol{C}\right)
\widetilde{\alpha}-g^{2}\boldsymbol{C}\widetilde{\beta}\right]  -\frac{\gamma
}{2}=0\\
\widetilde{a}g\left[  -\boldsymbol{C}\widetilde{\alpha}+\left\{
\boldsymbol{B+}g^{2}\boldsymbol{C}\right\}  \widetilde{\beta}\right]
-\frac{1}{2\gamma}=0\\
\widetilde{P}-2\pi g\widetilde{a}^{2}\left[  \widetilde{p}_{1}%
-\frac{\widetilde{a}^{2}}{3}\left(  \widetilde{\alpha}+\widetilde{\beta
}g^{2}\right)  \right]  =0
\end{array}
\right.  \label{dimlesssyst}%
\end{equation}
where we used $\widetilde{q}_{0}=\frac{\widetilde{Q}}{2\pi\widetilde{a%
}^{2}g}$. If, in place of the normal force $\widetilde{P}$, the normal indentation $\widetilde{\delta}$ is controlled, the last equation in (\ref{dimlesssyst}) is replaced by 

\begin{equation}
\widetilde{\delta}= \widetilde{b}  \left[  2\widetilde{p}_{1}\boldsymbol{K}%
-\widetilde{\alpha} \widetilde{a}^{2}\boldsymbol{B}-\widetilde{\beta} \widetilde{b}^{2}\boldsymbol{D}\right]
\label{ddless}%
\end{equation}

For a tangential displacement controlled experiment we recall that an
elliptical shear distribution as in (\ref{q}) produces a uniform tangential
displacement $\delta_{T}$ equal to \cite{Joh1985}%

\begin{equation}
\delta_{T}=\frac{Q}{\pi aE^{\ast}\left(  1-\nu\right)  }\left[
\boldsymbol{K-}\frac{\nu}{1-g^{2}}\left(  \boldsymbol{K-E}\right)  \right]
;\qquad b<a%
\end{equation}
where we used the identity $E^{\ast}=\frac{E}{1-\nu^{2}}$ and $q_{0}=\frac
{Q}{2\pi ab}.$ In dimensionless form $\widetilde{\delta}_{T}%
=\delta_{T}\xi^{2}/R$ gives%

\begin{equation}
\widetilde{\delta}_{T}=\frac{\widetilde{Q}}{\pi\widetilde{a}\left(
1-\nu\right)  }\left[  \boldsymbol{K-}\frac{\nu}{1-g^{2}}\left(
\boldsymbol{K-E}\right)  \right]  ;\qquad\widetilde{b}<\widetilde{a%
}\label{deltat}%
\end{equation}
so that $\widetilde{Q}$ may be replaced by $\widetilde{\delta}_{T}$ in
(\ref{dimlesssyst}).\\
Although the theoretical model has been derived with the hypothesis of having the tangential force $Q$ aligned with the minor axis (y direction in Fig. \ref{setup}b and c), it can be trivially rewritten with $Q$ aligned with the major axis.

\subsection{Mode-mixity function estimation}

As shown in Papangelo \& Ciavarella \cite{Pap2019} for the axisymmetric case, the model results are very sensitive to the exact choice of the phenomenological mode-mixity function. After testing the Literature models available, e.g. the models proposed by Hutchinson \& Suo \cite{Hutc1992}, we decided to extract the mode-mixity function from a calibration experiment. \\

Assume that for a given experimental set-up we know the geometry $\left(
R_{1},R_{2}\right)  ,$ the applied normal force $P$ (or indentation $\delta
$)$,$ and for each tangential force $Q$ the corresponding semi-axes of the
contact patch $\left(  a,b\right)  $. It is possible to estimate the
MMFs $f_{II}\left(  \psi_{2}\right)  $ and $f_{III}\left(  \psi_{3}\right)  $
by the following procedure. First, from (\ref{dimlesssyst}, eq. 3-4) one
obtains $\left(  \widetilde{\alpha},\widetilde{\beta}\right)  $%

\begin{align}
\widetilde{\alpha} &  =\frac{\gamma^{2}\boldsymbol{B}+\left(  1+\gamma
^{2}\right)  g^{2}\boldsymbol{C}}{2\widetilde{a}g\gamma\left[
g^{2}\boldsymbol{CD+B}\left(  \boldsymbol{C+D}\right)  \right]  }%
\label{alpha}\\
\widetilde{\beta} &  =\frac{\left(  1+\gamma^{2}\right)  \boldsymbol{C+D}%
}{2\widetilde{a}g\gamma\left[  g^{2}\boldsymbol{CD+B}\left(
\boldsymbol{C+D}\right)  \right]  }\label{beta}%
\end{align}
then, using (\ref{dimlesssyst}, eq. 5) one computes $\widetilde{p}_{1}$%

\begin{equation}
\widetilde{p}_{1}=\frac{\widetilde{P}}{2\pi g\widetilde{a}^{2}}%
+\frac{\widetilde{a}^{2}}{3}\left(  \widetilde{\alpha}+\widetilde{\beta
}g^{2}\right)  \label{p1}%
\end{equation}
hence finally from (\ref{dimlesssyst}, eq. 1-2) one obtains%

\begin{align}
f_{II,\text{exp}}\left(  \psi_{2}\right)   &  =\frac{\pi g\widetilde{a}%
}{2}\left[  \left(  \widetilde{\beta}g^{2}\widetilde{a}^{2}-\widetilde
{p}_{1}\right)  ^{2}+\left(  \frac{\widetilde{Q}}{2\pi\widetilde{a}^{2}%
g}\right)  ^{2}\right]  \label{fIIexp}\\
f_{III,\text{exp}}\left(  \psi_{3}\right)   &  =\frac{\pi\widetilde{a}}%
{2}\left[  \left(  \widetilde{\alpha}\widetilde{a}^{2}-\widetilde{p}%
_{1}\right)  ^{2}+\frac{1}{1-\nu}\left(  \frac{\widetilde{Q}}{2\pi\widetilde{a}^{2}%
g}\right)  ^{2}\right]  \label{fIIIexp}%
\end{align}
The corresponding phase angles will be for mode I-II interaction%

\begin{equation}
\psi_{2}=\arctan\left(  \frac{K_{II}}{K_{I}}\right)  =\arctan\left(
\frac{\widetilde{Q}}{2\pi\widetilde{a}^{2}g\left(  \widetilde{\beta
}\widetilde{a}^{2}g^{2}-\widetilde{p}_{1}\right)  }\right)
\end{equation}
and for mode I-III interaction%

\begin{equation}
\psi_{3}=\arctan\left(  \frac{K_{III}}{K_{I}}\right)  =\arctan\left(
\frac{\widetilde{Q}}{2\pi\widetilde{a}^{2}g\left(  \widetilde{\alpha
}\widetilde{a}^{2}-\widetilde{p}_{1}\right)  }\right)
\end{equation}

\section{Comparison with Experimental results}

\subsection{Determining the mode-mixity function}

Let us consider the experimental data discussed in the companion Letter \cite{Letter} and in Sahli et al. \cite{Sahli}. The experimental set-up is composed of a cantilever which sustains a glass substrate which is pressed against a
PDMS sphere of radius $R$ and then sheared (see Fig. \ref{setup}c). A camera was used to track the contact area evolution while a force
cell simultaneously measured the tangential force applied.  The experimental results reported by Sahli et al. \cite{Sahli} and further analyzed in the companion
Letter \cite{Letter} are provided for the following set of normal forces $P=\left[
0.27,0.55,0.82,1.10,1.37,1.65,1.92,2.12\right] $ $N$ which span one order of magnitude and for the following sphere radius 
$R=9.42$ mm.

To estimate the MMFs the aforementioned
procedure was used, i.e. the equations (\ref{alpha},\ref{beta},\ref{p1},\ref{fIIexp}%
,\ref{fIIIexp}), for the arbitrarily selected data corresponding to
the case $P=0.55$ N\footnote{Similar results can be obtained selecting the set
of data corresponding to a different normal force.}. For the PDMS/glass interfaces we used the following material properties (see \cite{Sahli} and their {\it Supporting Information}) 

\begin{equation}%
\begin{array}
[c]{c}%
G_{Ic}=27\text{ mJ/m}^{2}%
;\text{\quad\quad}E=1.88\text{ MPa};\\
\nu=0.5;\text{\quad\quad}\sigma=0.41\text{ MPa};
\end{array}
\label{param}%
\end{equation}

\begin{figure}[ptb]
\begin{center}
\includegraphics[width=3.5in]{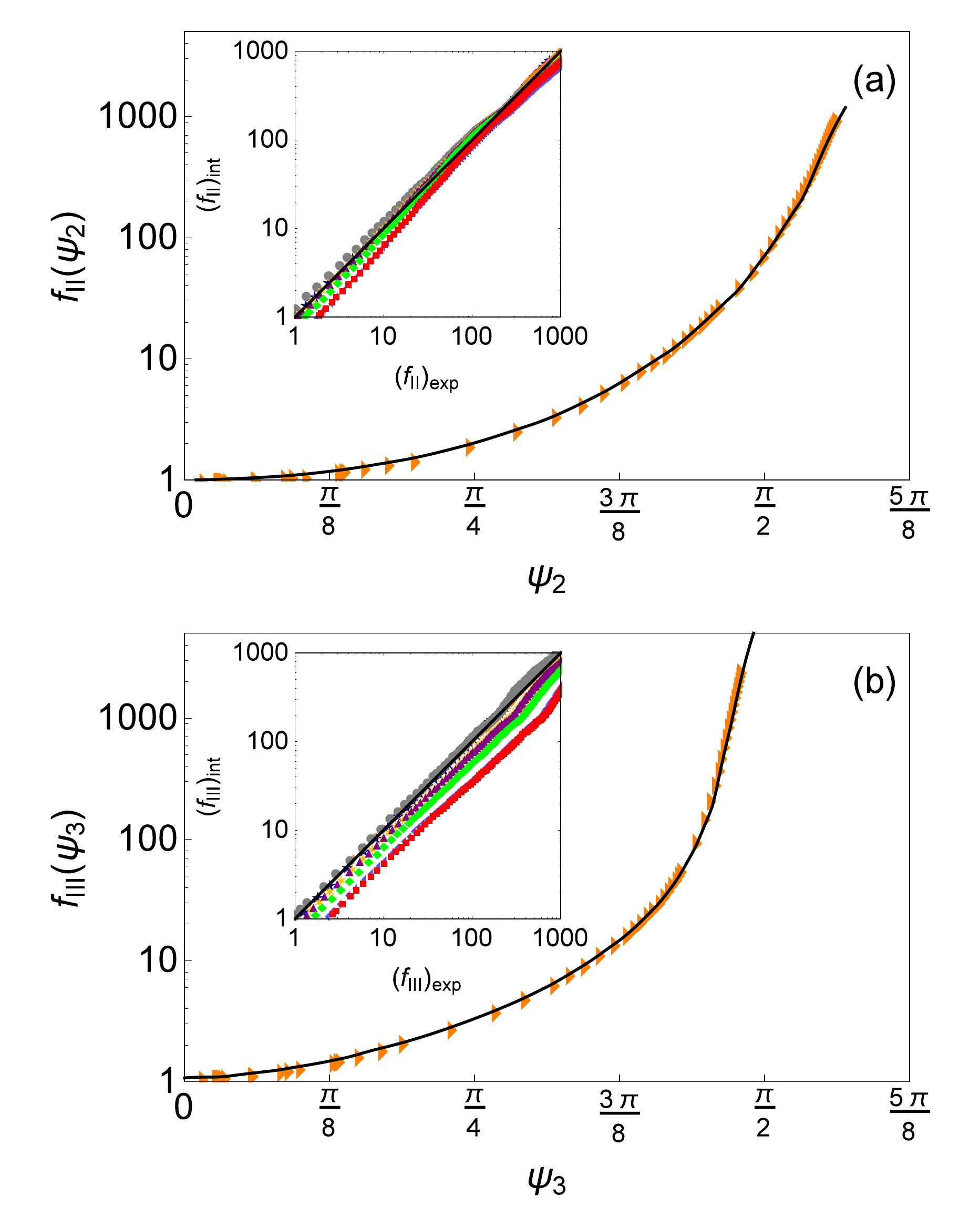}
\caption{Orange triangles: mode-mixity functions (a) $f_{II}\left(  \psi_2\right)$ (respectively (b) $f_{III}\left(  \psi_3\right)$)  estimated from eq. (\ref{fIIexp}) (respectively (\ref{fIIIexp})) from the experimental data of case $P=0.55$ N. Solid lines: interpolation of the curves used in the comparison with other experimental data available in the companion Letter \cite{Letter}. \rev{Insets: interpolated vs experimental MMFs for all experiments. Solid lines: equality lines. Markers for experimental results. Blue stars, orange right-triangles, gray circles, yellow stars, purple up-triangles, green diamonds, violet left-triangles, red squares respectively for
$P=\left[ 0.27,0.55,0.82,1.10,1.37,1.65,1.92,2.12\right]  N$.}}%
\label{Figpsi}%
\end{center}
\end{figure}

\begin{figure}[ptb]
\begin{center}
\includegraphics[width=3.5in]{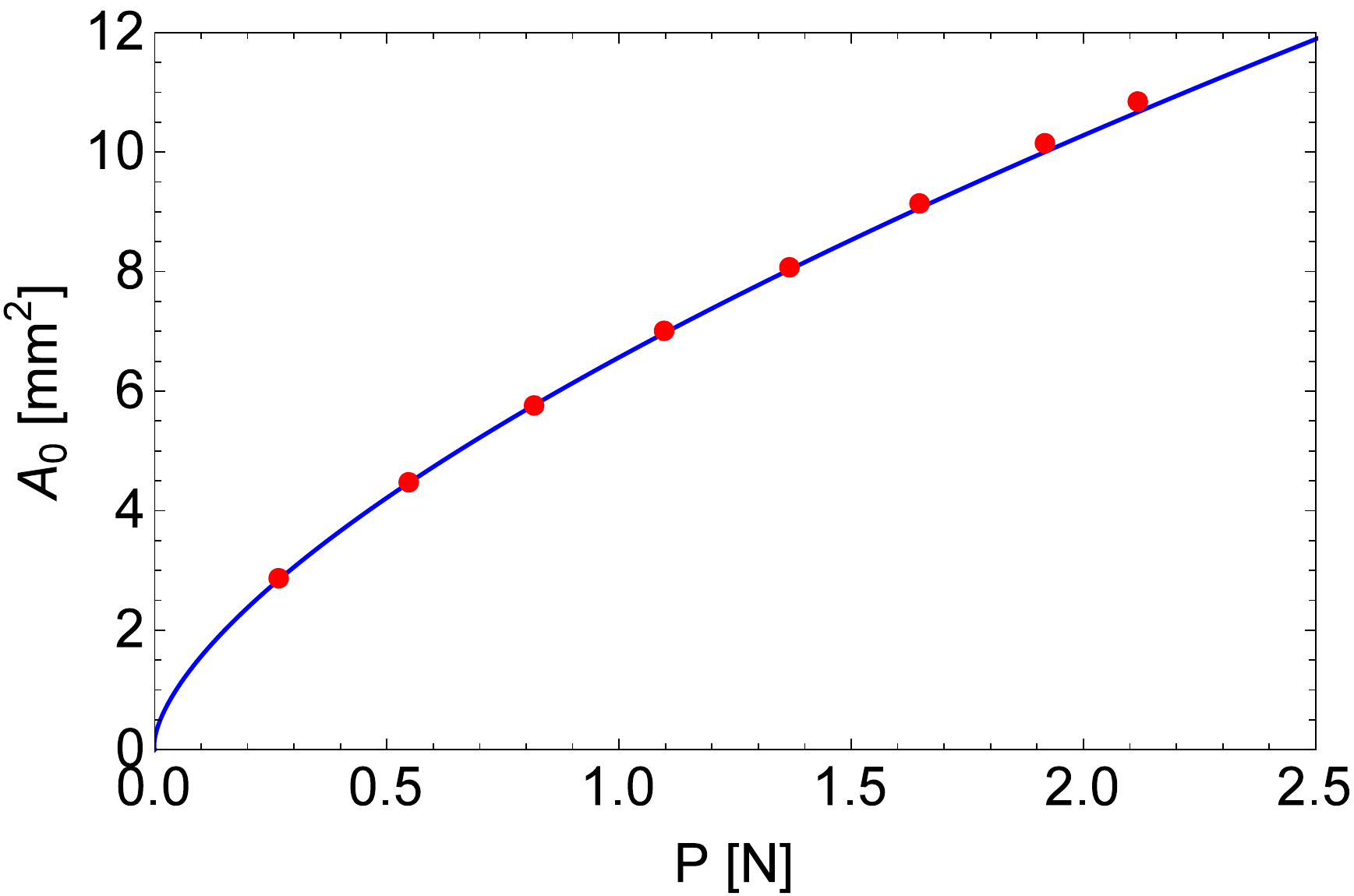}
\caption{Contact area under null tangential force $A_0$ vs normal force $P$. Solid line: JKR model with $G_{Ic}=27\text{ mJ/m}^{2}, E=1.88\text{ MPa}, \nu=0.5$ and $R=9.42$ mm. Red dots: experimental data under null tangential force.}%
\label{FigJKR}%
\end{center}
\end{figure}

\noindent{where} $\sigma$ is the best fitted average shear strength of the interface (see Fig. \ref{AQ}) and $E$ was obtained from the control experiment\footnote{For the control experiment with $P=0.55$ N, under zero tangential force, a contact area $A_0\simeq4.48$ mm$^2$ was measured. Using the JKR relation $P=\frac{4E^{\ast}}{3R}{\left( \frac{A_0}{\pi}\right)}^{3/2}-\sqrt{8\pi^{-1/2} E^{\ast}{A_0}^{3/2}G_{Ic}}$ with $G_{Ic}=27\text{ mJ/m}^{2}$, $\nu=0.5$ and $R=9.42$ mm one gets $E\simeq1.88$ MPa.} 
with $P=0.55$ N. Figure (\ref{Figpsi}a) shows the experimental data (orange
triangles) and the interpolated (black solid line) MMF
$f_{II}\left(  \psi_{2}\right)  $ as a function of the phase angle $\psi_{2}$. {$f_{II}\left(
\psi_{2}\right)  $ can be well approximated by $\log[f_{II}\left(  \psi
_{2}\right)  ]=a_{2}\psi_{2}^{2}+b_{2}\psi_{2}^{n_{2}}$, where the
coefficients are $\left(  a_{2},b_{2},n_{2}\right)  =\left(  1.18,5.67\ast
10^{-2},7.05\right)  .$ 
To obtain a\ better fit, the data were interpolated in log-linear form, i.e.
$\left(  \psi_{2},\log[f_{II}\left(  \psi_{2}\right)  ]\right)  $, which
allows to catch the MMF across all scales. The inset shows the interpolated
mode-II MMF versus the one evaluated from all the set of experimental data
available, which do
collapse over 3 orders of magnitude of $f_{II}$. With the same procedure, $f_{III}\left(
\psi_{3}\right)  $ has been interpolated from the experimental data using
solely the set of data corresponding to the case $P=0.55$ $N$ (Fig.
\ref{Figpsi}b)$.$ $f_{III}\left(  \psi_{3}\right)  $ can be well
approximated by $\log[f_{III}\left(  \psi_{3}\right)  ]=a_{3}\psi_{3}%
^{2}+b_{3}\psi_{3}^{n_{3}}$, where the coefficients are $\left(  a_{3}%
,b_{3},n_{3}\right)  =\left(  1.87,6.73\ast10^{-3},15.20\right)  .$} The inset shows that the complete set of experimental data
align along the main diagonal, nevertheless the data referring to the higher
normal forces, i.e. $P\approx\left[  1.65,1.92,2.12\right]  $ $N$, appear to be
shifted by a factor $\approx2$ also for vanishing tangential forces $\left(  f\left(  \psi\right)
\simeq1\right)  ,$ which indicates a small deviation in the original JKR fit (see the 3 rightmost points in Fig. \ref{FigJKR}).
\rev{It is worth noting that the normal force is varying by one order of magnitude
in the same set of experiments, hence some nonlinear effects (probably due to
stiffening in the material) may have arisen which make the JKR fit not
perfect.} Figure \ref{FigJKR} shows the JKR curve (black solid line) obtained
with the parameters reported in the companion Letter \cite{Letter} and by Sahli et al. \cite{Sahli}
(see (\ref{param})) and for each normal force the contact area under null shear
force (red dots). It can be observed that the deviations from JKR\ are very small.%


\subsection{Decay of contact area}

\begin{figure}
[ptb]
\begin{center}
\includegraphics[width=3.5in]{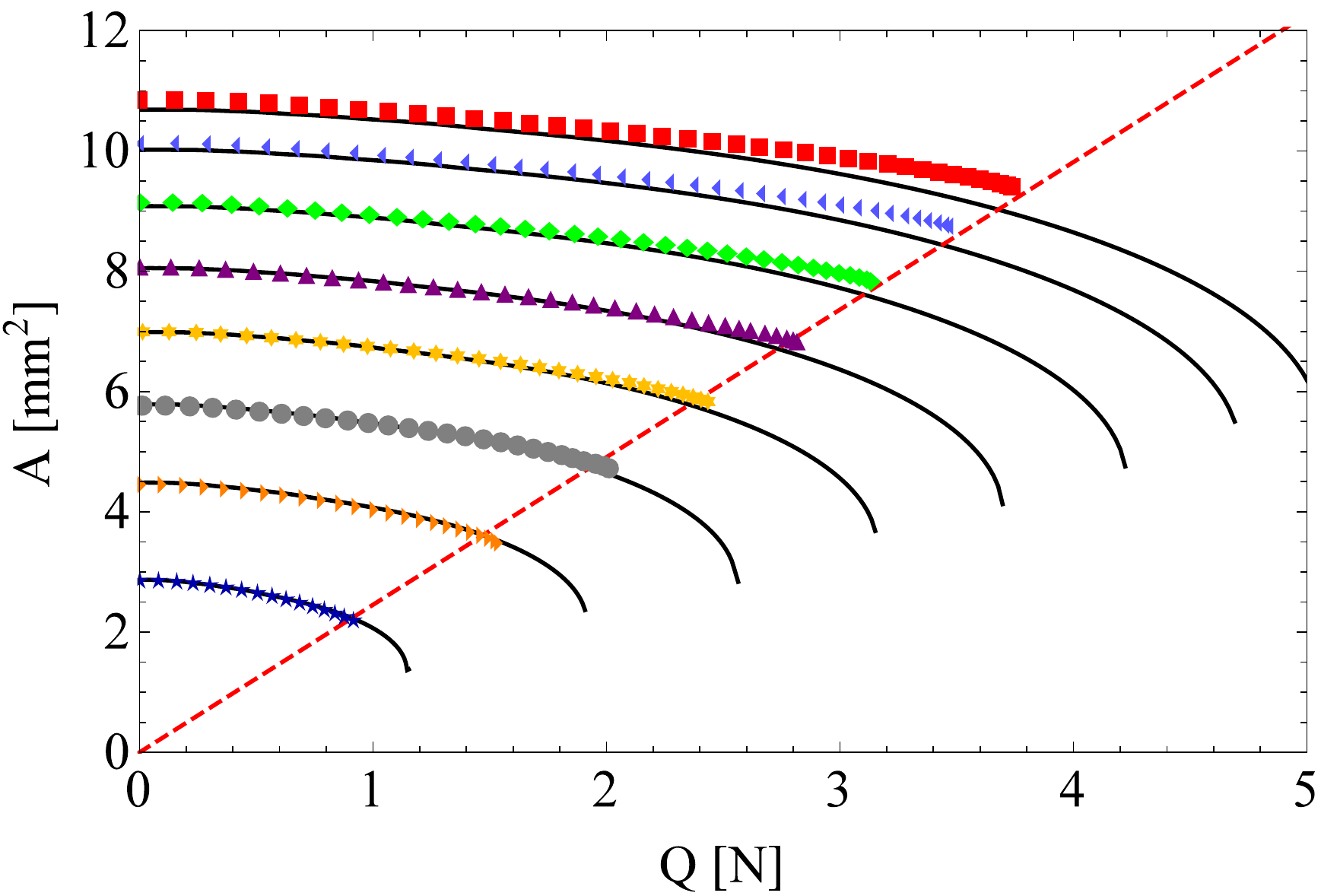}
\caption{Contact area $A$ as a function of the tangential force $Q$ for
different normal forces $P=\left[
0.27,0.55,0.82,1.10,1.37,1.65,1.92,2.12\right]$ N and $R_{e}=R=9.42$ mm. The markers indicate the
experimental measurements, the solid black lines show the model prediction
while the dashed red line indicates the full sliding criterion $Q_{s}=\sigma\ast
A$ with $\sigma=0.41$ MPa. }%
\label{AQ}%
\end{center}
\end{figure}

\begin{figure}
[ptb]
\begin{center}
\includegraphics[width=3.5in]{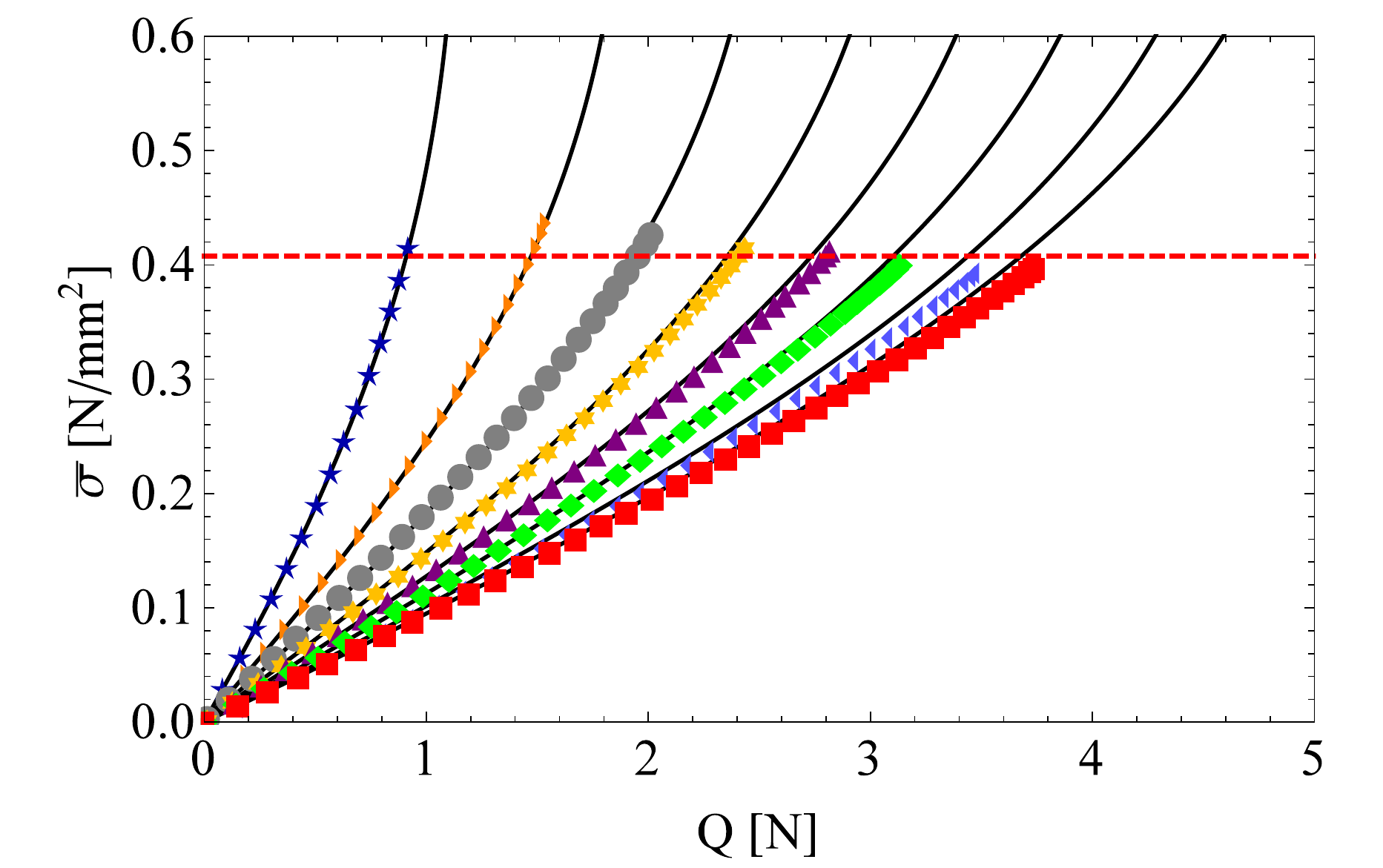}%
\caption{Mean shear stress at the interface $\overline{\sigma}=Q/A$ according to the experimental
results from Sahli et al. \cite{Sahli} (markers) and to the proposed model (solid black lines). The solid lines are drawn for
$P=\left[  0.27,0.55,0.82,1.10,1.37,1.65,1.92,2.12\right]  N$ and $R_{e}=R=9.42$ mm. The red
dashed line marks the full sliding points at $\overline{\sigma}=\sigma=0.41$ MPa. }%
\label{Figdelta}%
\end{center}
\end{figure}

In this section the results obtained solving the system of equations
(\ref{dimlesssyst}) are presented, where the unknown MMFs $f_{II}\left(
\psi_{2}\right)  $ and $f_{III}\left(  \psi_{3}\right)  $ have been
substituted by the one estimated in the previous section using only the data
set for $P=0.55$ N. Fig. \ref{AQ} shows the contact area evolution as a
function of the tangential force for the complete set of experimental data from Sahli et al. \cite{Sahli} with $R_{e}=R=9.42$ mm (PDMS sphere/glass substrate contact). \rev{The
markers indicate the experimental results obtained for each normal force,
while the black solid lines are for the proposed model, that proves to be in
very good agreement with all the observations. Small deviations appear for the
heavier normal forces as was already found and discussed in the previous section.} The
dashed red line shows the full sliding threshold according to the criterion
$Q_{s}=\sigma\ast A$, as proposed by Sahli et al. \cite{Sahli} and
Mergel et al. \cite{Mergel}. Figure \ref{Figdelta} favorably
compares the mean shear stress at the interface $\overline{\sigma}=Q/A$
according to the experimental results (markers) and to the proposed model
(solid black lines), where the red dashed lines marks the boundary of the full
sliding region, i.e. $\overline{\sigma}=\sigma=0.41$ $MPa$.%

%

\bigskip

\section{Contact shearing along the major/minor semi-axes}

\begin{figure}
[ptb]
\begin{center}
\includegraphics[width=3.5in]{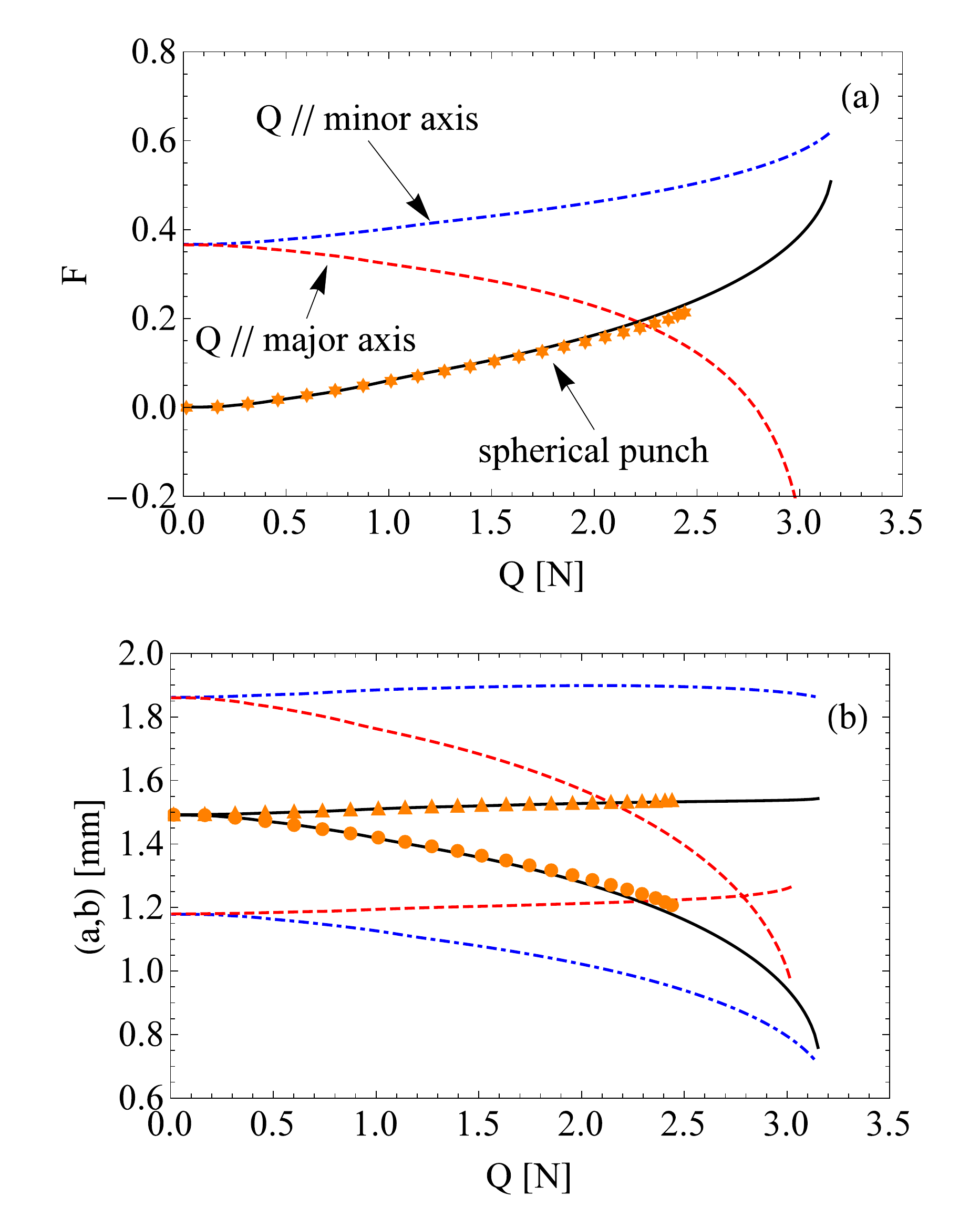}%
\caption{(a) Ellipticity $F=1-b/a$ versus the tangential force $Q$ as obtained
experimentally for $P=1.10$ N, $R_{e}=R=9.42$ mm (orange stars) and as obtained from the
model (solid black line). Prediction of the ellipticity evolution
for a non-axisymmetric punch with $R_{2}/R_{1}=1/2$ ($R_{e}=9.42$ mm) loaded along its major
(dashed red line) or minor (blue dotdashed line) axis. (b) Evolution of the
the semi-axes $\left(a,b\right)  ,$ with $a>b$ at $Q=0$ N. Symbols and lines
as in panel (a).}%
\label{gab}%
\end{center}
\end{figure}

Let us compare the model predictions with the experimental results
in terms of evolution of the ellipticity (or flattening) $F=1-b/a$. For this comparison, the set of
experimental data for $P=1.10$ N and $R_{e}=R=9.42$ mm has been chosen. In Fig. \ref{gab}a the
ellipticity is plotted against the tangential force $Q$: the
experimental data are plotted with orange stars, while the model prediction is
shown as a black solid line. The same set of data is plotted in Fig.
\ref{gab}b in terms of evolution of the semi-axes $\left(  a,b\right)
$. Notice that the contact area shrinks drastically along the direction aligned with the tangential force,  semi-axis ``$b$'', while the perpendicular axis ``$a$'' remains mostly unaffected by the tangential force. This is in agreement with the observation that the interfacial toughness under the mode combination I-III was found greater than under mode I-II combination (compare $f_{II}\left(  \psi_2\right)  $ and $f_{III}\left(  \psi_3\right)  $ in Fig. \ref{Figpsi}). The predictions are in excellent agreement with the experimental results of \cite{Letter}.

We then investigated the indentation of a non-axisymmetric punch with
$R_{e}=R=9.42$ mm and $R_{2}/R_{1}=1/2$, so as in the typical rough contacts
according to Greenwood \cite{G2006}, being all the other parameters unchanged. For
the latter case no experimental data are available to compare with, thus only
the model predictions are presented. Figure \ref{gab}a shows the evolution of
the ellipticity when the punch is loaded along its major (red dashed line) and
minor (blue dotdashed line) axis. The same results are plotted in terms of
semi-axes evolution in Fig. \ref{gab}b. Notice that after shearing, the contact patch shapes are
strongly different among the three cases we have analyzed, i.e. axisymmetric punch, and non-axisymmetric punch loaded along the major or minor axis. Indeed the axis under mode
II loading tends to shrinks much more rapidly with respect to the axis under
mode III loading. Hence the punch loaded along its major axis shrinks towards
a more circular shape, i.e. the ellipticity decreases, and eventually becomes
negative as due to the shearing force, we obtain $a<b$. On the contrary, loading along the minor axis produces a contact
patch with increasing ellipticity while $Q$ is increased.

\begin{figure}
[ptb]
\begin{center}
\includegraphics[width=3.5in]{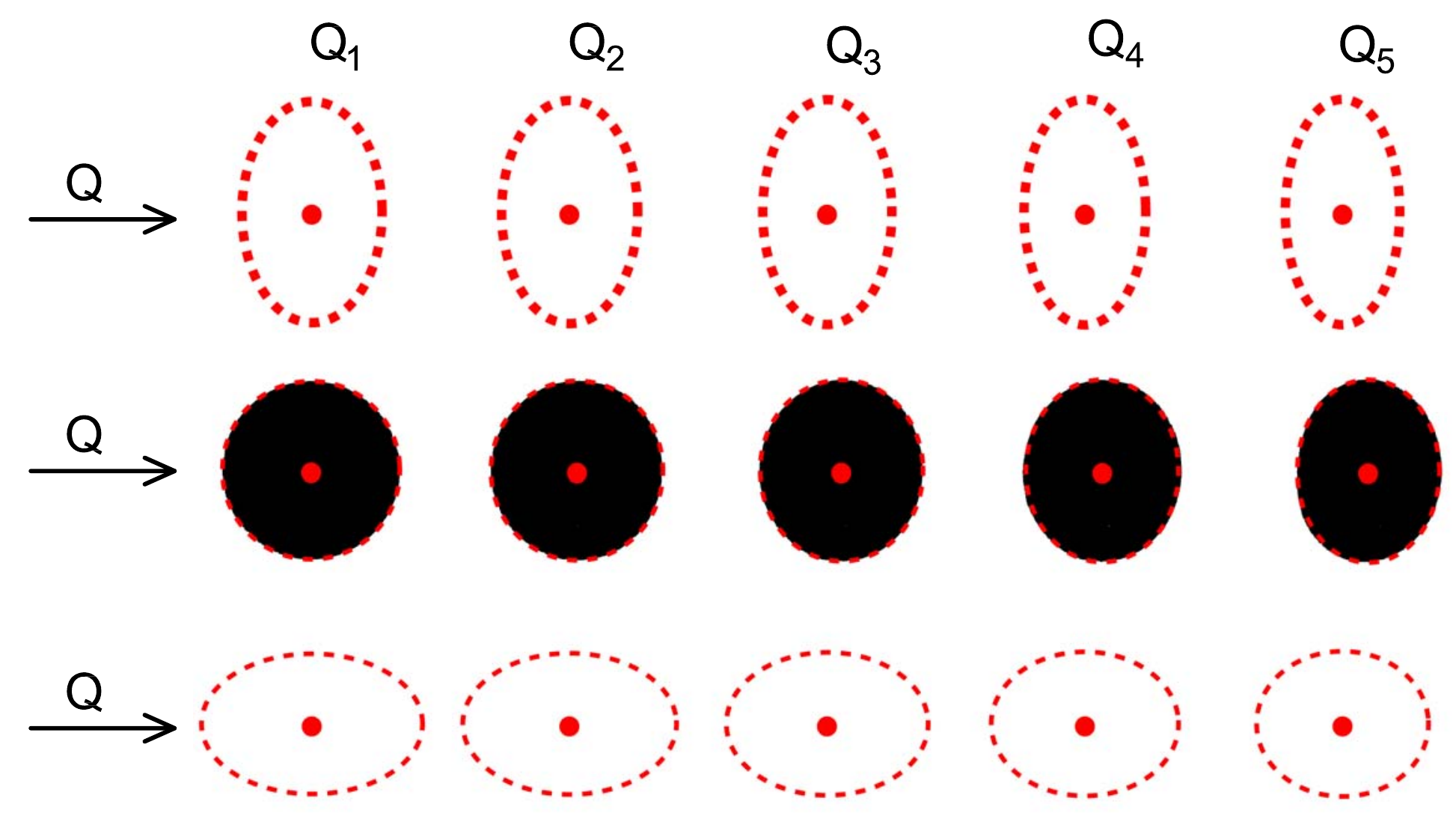}
\caption{Evolution of the contact patches according to the proposed model (red
dashed line) for: $R_{2}/R_{1}=1/2,R_{e}=R$ loaded respectively along the minor
(top row) and major (bottom row) axis and for $R_{1}=R_{2}=R$ (middle row). For the
axisymmetric case (middle row) experimental contact patches are plotted in black. The tangential forces range from $Q_{1}$ to $Q_{5}$, respectively
$[0.04,0.77,1.42,1.98,2.38]$ N. For all the snapshots $P=1.10$ N.}%
\label{snap}%
\end{center}
\end{figure}

The theoretical model is based on the assumption that the contact area shrinks
in an elliptical fashion while the contact is sheared. In Fig. \ref{snap} we
check this assumption comparing with actual experimental snapshots of the
contact area (same data used for Fig. \ref{gab}) taken for 5 tangential forces,
from $Q_{1}$ to $Q_{5}$ respectively $[0.04,0.77,1.42,1.98,2.38]$ N. The
results are reported for $R_{e}=R$ and respectively $R_{2}/R_{1}=1$ (middle
row) and $R_{2}/R_{1}=1/2$ (top and bottom row) where the shearing force is
aligned with the minor (top row) and major (bottom row) axis. The evolution of
the contact patches according to the proposed model is shown as a red dashed
line (all rows) while the experimental contact patches are plotted as
a black patch (middle row). The agreement between experimental results and
model prediction is excellent for the axisymmetric punch, while we can provide only predictions for $R_{2}/R_{1}=1/2$ as experimental data are missing.%

\begin{figure}
[ptb]
\begin{center}
\includegraphics[width=3.5in]{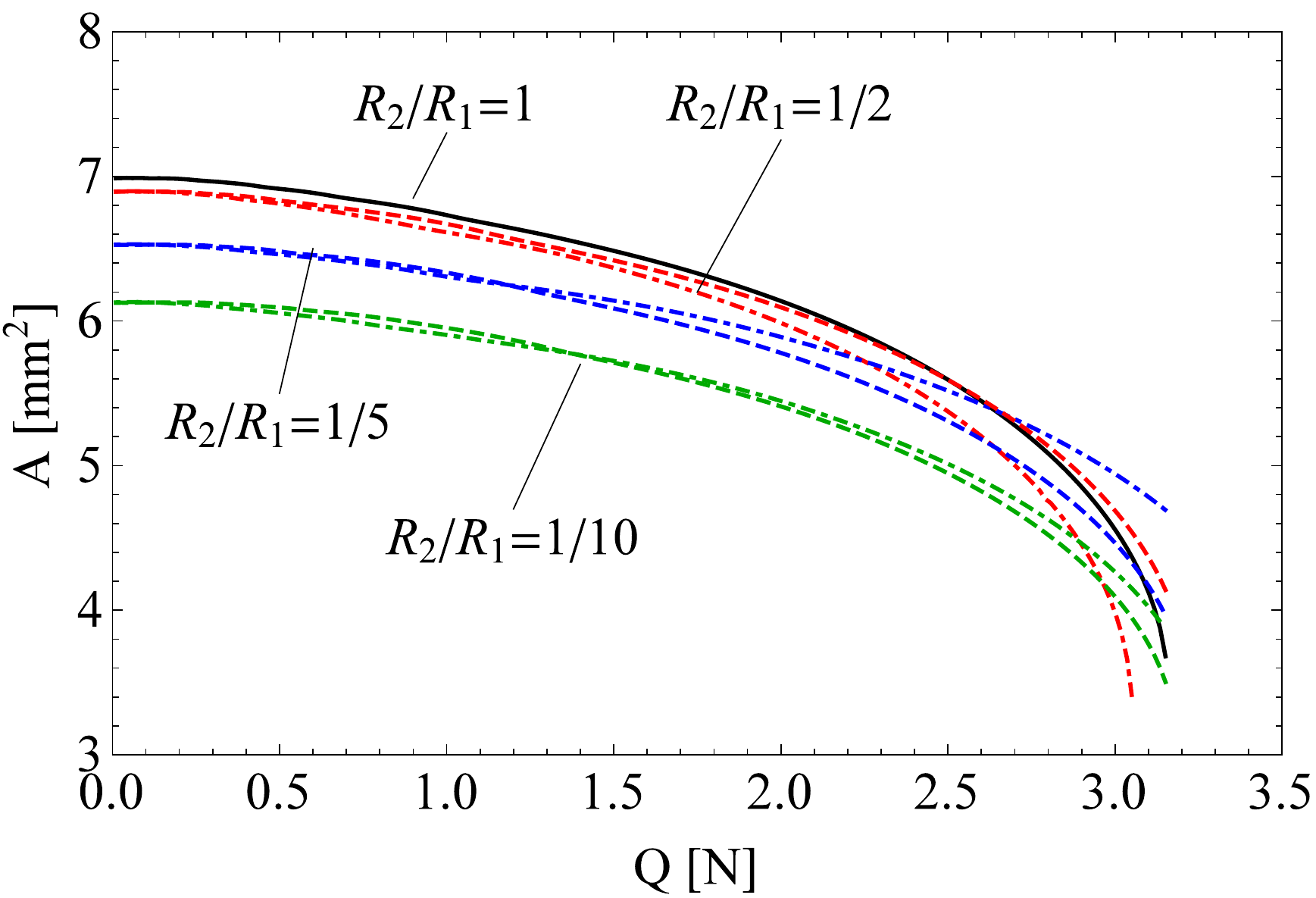}
\caption{Evolution of the contact area with the tangential force reported
for $R_{e}=R=9.42$ mm, $P=1.10$ N, $R_{2}/R_{1}=\left[
1,1/2,1/5,1/10\right]  $ respectively solid black
line, red, blue and green lines. Predictions have been made for both $Q$
aligned with the major (dotdashed lines) and minor (dashed lines) axis.}%
\label{AQAgamma}%
\end{center}
\end{figure}

Finally, we further explore the effect of the initial geometry, $R_{2}/R_{1}$
ratio, on the contact area decay. In Fig. \ref{AQAgamma} the evolution of
contact area with the tangential force is reported for $R_{e}=R=9.42$ mm, $P=1.10$ N, $R_{2}/R_{1}=\left[
1,1/2,1/5,1/10\right]  $ respectively solid black line, red, blue and green
lines. Predictions have been made for both $Q$ aligned with the major
(dotdashed lines) and minor (dashed lines) axis. One concludes that the
contact shapes are affected by both the punch geometry and the direction of shear with respect to the ellipse orientation. Inspection of Fig. \ref{AQAgamma} reveals that in terms of overall contact area decay for increasing shear force $Q$, changing the ratio $R_{2}/R_{1}$ from $1$ to $1/10$ will produce a reduction of the overall contact area of
the order of $10-15\%$ for both $Q$ aligned along the major
(dotdashed line) or minor (dashed) axis.%

\section{Scaling law for area decay}

In their paper, Sahli et al \cite{Sahli} showed that for smooth spheres a
quadratic form $\ A(Q)=A_{0}-\alpha_{A}Q^{2}$ well captures the decay of
contact area with tangential force, where $A_{0}$ is the contact area for $Q=0$
N and $\alpha_{A}$ is a fitting coefficient. Interestingly, they found that
$\alpha_{A}$ shows a power law scaling with $A_{0}$ with exponent $-3/2$ over
4 orders of magnitude which comprises data from interfacial microjunctions
(rough contacts) and data from smooth spheres. Literature LEFM axisymmetric models for smooth spheres (\cite{Ciavafacta},
\cite{Pap2019}) have found a similar but not equal exponent, i.e. $-5/4$.

\begin{figure}
[ptb]
\begin{center}
\includegraphics[width=3.5in]{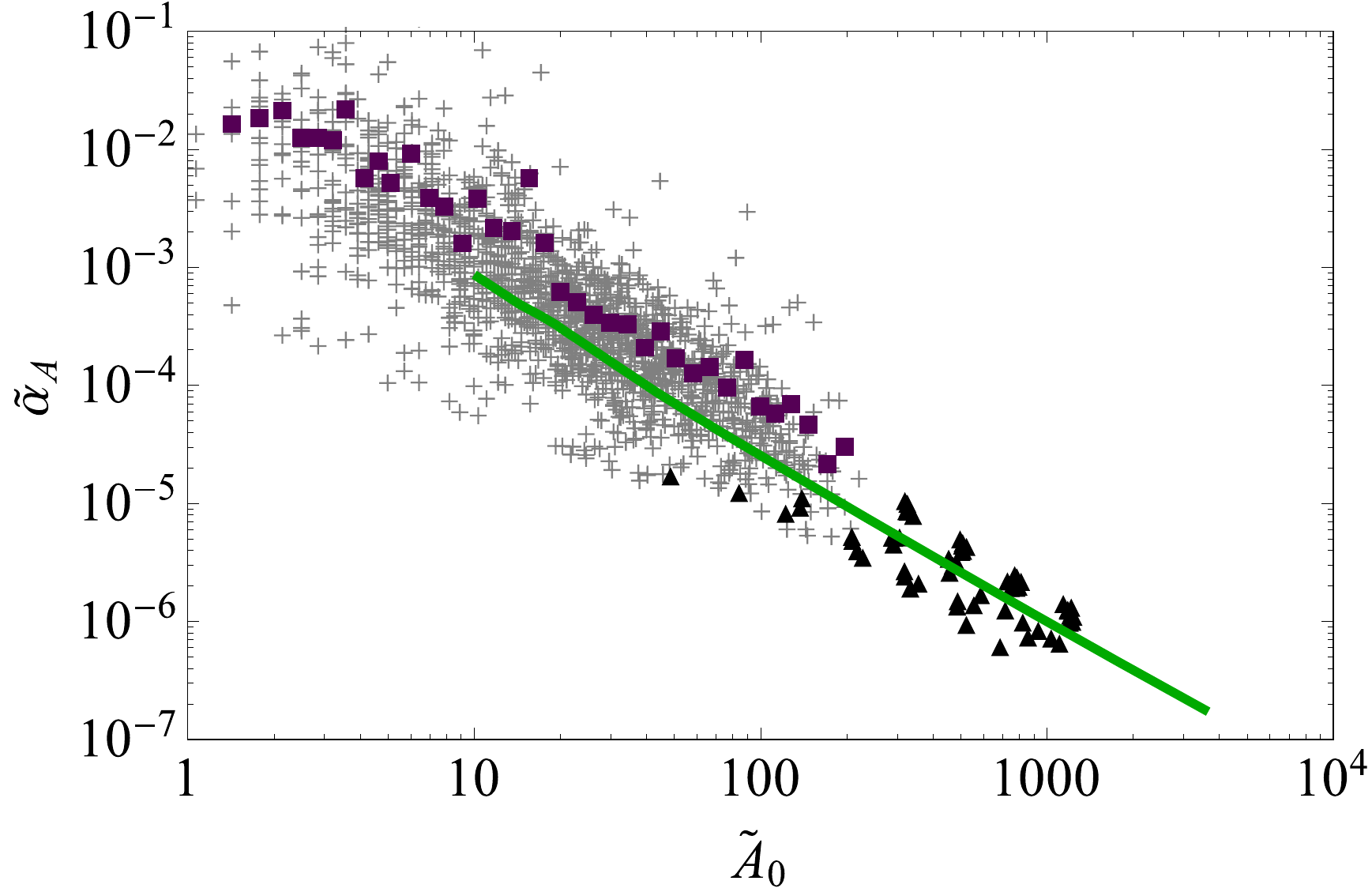}%
\caption{Greeen solid line: coefficient $\widetilde{\alpha}_{A}$ obtained fitting the numerical data obtained by the proposed
elliptical model with a power law function $\widetilde{A}(Q)=\widetilde{A}_{0}%
-\widetilde{\alpha}_{A}\widetilde{Q}^{2}$ as proposed by Sahli et al. \cite{Sahli} as a function of $\widetilde{A_0}$. 
Symbols represent the experimental data reported in \cite{Sahli}: triangles for smooth sphere with radii $R=[7.06, 9.42, 24.81]$ mm, crosses refer to the experimental data obtained for microjunctions (rough contact) with average values indicated by purple squares.}%
\label{Figalpha}%
\end{center}
\end{figure}

Here we investigate which scaling law would arise from the present elliptical model and compare with experimental results. We defined a set of normal forces ranging from $1$ mN to $10$ N and, using the model, obtained the area vs
tangential force curves up to full sliding, i.e. truncating them at
$Q_{s}=\sigma\ast A$. We used the same material properties (\ref{param}) and geometry parameters $R_{e}=R=9.42$ mm adopted in the previous analyses. The resulting curves were fitted with a quadratic area decay law that in dimensionless form reads $\widetilde{A}(\widetilde{Q})=\widetilde{A}_{0}-\widetilde{\alpha}_{{A}}\widetilde{Q}^{2}$, being $\widetilde{A}=A(\xi/{R_{e}})^2$ and $\widetilde{\alpha}_{A}= \alpha_{A} \left(\xi G_{Ic}\right)^2$. To this end we needed to estimate the mean microjunction radius. We consider the results of a single rough contact experiment from Ref.~\cite{Sahli}:, PDMS/glass contact, under a normal force of $P=6.40$ N for which $514$ microcontacts with an initial area larger than $2\ast10^{-9}$ m$^{2}$ were found and tracked. This results in an average force for each microjunction equal to $P_{i}=6.4/514\simeq1.2\times10^{-2}$ N. From the distribution of microjunctions contact areas we derived the characteristic dimension of the microjunction $a_{i}=\sqrt{A_{i}/\pi}$ and computed the mean contact radius $\overline{a}_{i}=0.235$ mm. Using the JKR model with known $P_{i},a_{i}$ and material properties $G_{Ic}=27\text{ mJ/m}^{2}, E=1.88\text{ MPa}, \nu=0.5$ we estimated the mean radius of curvature $R_{micro}\approx2.6$ mm. In Fig. \ref{Figalpha},$\widetilde{\alpha}_{A}$ is shown as a function of $\widetilde{A}_{0}$ (green solid line), with superimposed the experimental data obtained for smooth spheres (black triangles) and microjunctions (raw data: gray crosses, averaged data: purple squares). The agreement between the model and the experimental results is very good over more than 2 orders of magnitude in $\widetilde{A}_{0}$, but cannot be assessed in the range $1 < \widetilde{A}_{0} < 10$. Indeed JKR theory predicts, under force control, that the smallest stable contact spot is $\widetilde{A}_{min,JKR} =\pi ( \frac{9 \pi}{8} )^{2/3}\simeq7.3$.

\begin{figure}
[ptb]
\begin{center}
\includegraphics[width=3.5in]{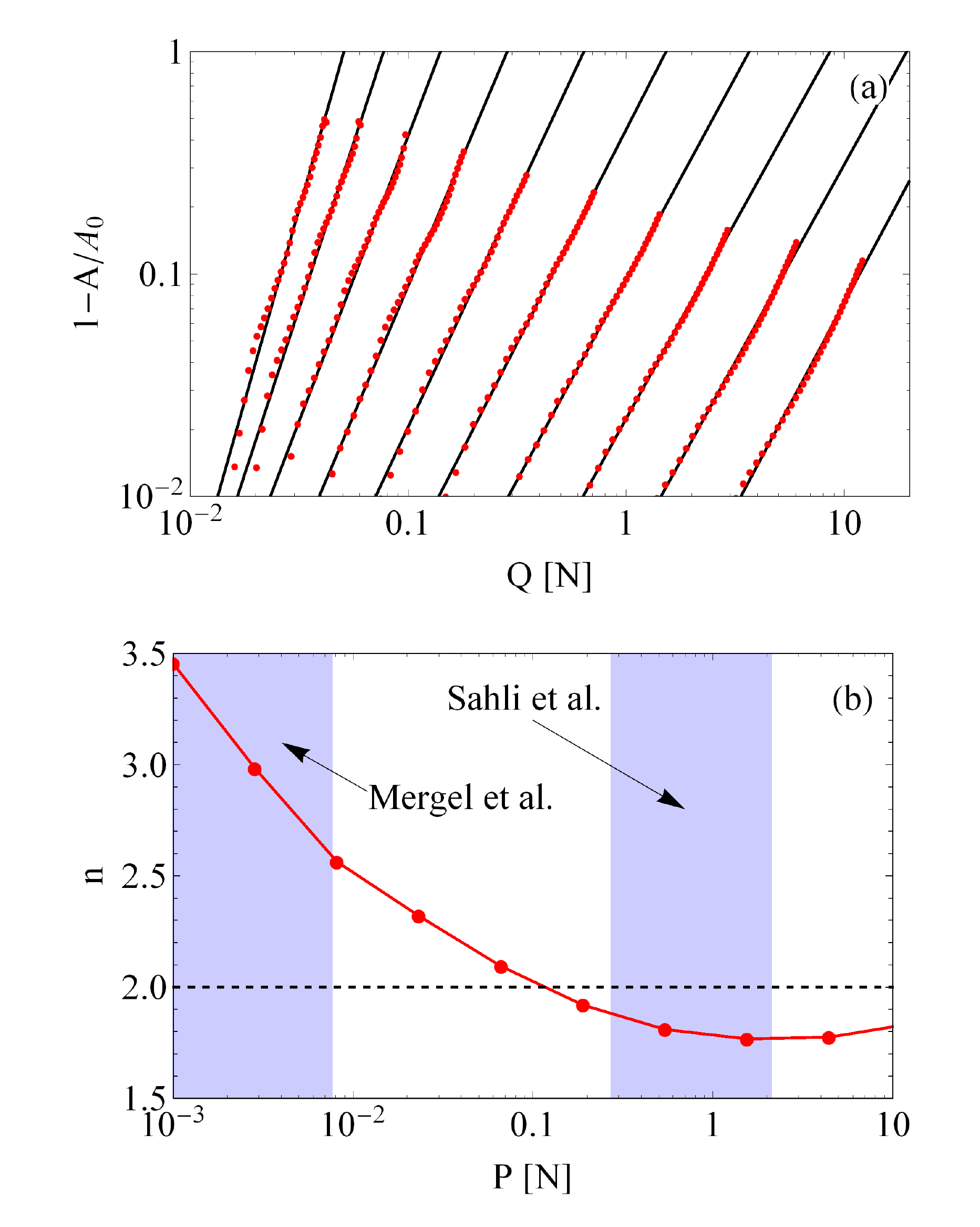}%
\caption{(a) Best fit of the form $1-A(Q)/A_0 \propto Q^{n}$ applied to the numerical data obtained by the proposed elliptical model for a range of normal forces ranging from $1$ mN to $10$ N.
Red dots represent the numerical data, while the solid black lines stand for the best fitted power law. The exponent $n$ is reported in panel (b) as a function of the normal force. Shaded areas indicate the regions where Sahli et al. \cite{Sahli} and Mergel et al. \cite{Mergel} data lie.}%
\label{Fign}%
\end{center}
\end{figure}

Discrepancies may arise at too small contact areas as the decay law may not be strictly quadratic anymore as indeed recent investigations seem to suggest \cite{Pap2019,Mergel}. We reconsidered the obtained area-force curves and fitted them using a power law function with form $A(Q)=A_{0}-c_{1}Q^{n},$ from which the best fit exponent $n$ has been obtained.  Figure \ref{Fign}a shows the quantity $1-A(Q)/A_{0}$ as a function of $Q$ in a log-log plot. The red
dots represent the points obtained using the elliptical model while the black solid lines are the best fitted power law functions obtained varying the normal force $P$ over 4 orders of magnitude. One easily recognizes that the lighter the normal force the steeper gets the power law function suggesting that a unique exponent is unlikely to best fit all the curves. In Fig. \ref{Fign}b $n$ is reported as a function of the normal force $P$ (solid curve). The shaded areas indicate the range of
normal forces used in the experiments by Sahli et al. \cite{Sahli} and Mergel
et al. \cite{Mergel}. Inspection of the graph reveals that for the experiments by Sahli et al. \cite{Sahli} normal forces are of the order of $1$ N and the contact area decay in the model is well fitted by a quadratic
power law $(n\approx1.8-1.9)$, while for lighter normal forces of the order
of $10^{-3}-10^{-2}$ N, as in \cite{Mergel}, a larger exponent is found
$n\approx3\pm0.5$. 


\section{Conclusions}

We have introduced the first non-axisymmetric model which successfully predicts
the anisotropic shearing of the contact area under adhesive conditions\ due to
tangential force. The model has been validated against several experimental
data from Sahli et al. \cite{Sahli} and included in the companion Letter \cite{Letter} and essentially
an excellent agreement is found. The model is based on LEFM and has been
inspired by the seminal work of JG, which has been extended to accomplish
tangential loading of the contact area. Using our elliptical model we have made predictions of contact area
evolution for non-axisymmetric punches. The results show that the effect of
differing principal radii of curvature strongly affects the evolution of the
contact shape. This may reveal to be a fundamental phenomenon in the development of contact patch anisotropy in rough contact under shear, where asperities are expected to be mildly elliptical \cite{G2006}. We have also shown that in terms of overall variation of contact area a reduction of $10-15\%$ can be expected varying $R_{2}/R_{1}$ from $1$ to $1/10$. Deviations from this behaviour may be expected due to the interactions between asperities, but this is out of the scope of the present paper.\\

\section*{Acknowledgements}

A.P. is thankful to the DFG (German Research Foundation) for funding the 
project PA 3303/1-1. M.C. is supported by the Italian Ministry of Education, University and Research (MIUR) under the “Departments of Excellence” grant L.232/2016. This work was supported by LABEX MANUTECH-SISE (ANR-10-LABX-0075) of Universit\'{e} de Lyon,
within the program Investissements d'Avenir (ANR-11-IDEX-0007) operated by the French National Research
Agency (ANR). It received funding from the People Program (Marie Curie Actions) of the European Union's
Seventh Framework Program (FP7/2007-2013) under Research Executive Agency Grant Agreement PCIG-GA-2011-303871. We are indebted to Institut Carnot Ing\'{e}nierie@Lyon for support and funding.

\section*{Author contribution statement}

A.P. and M.C. conceived the theoretical model and wrote the work. A.P. created the figures. J.S., R.S and G.P. provided the experimental data. All the authors revised the work up to its final form.

\end{document}